\documentclass[aps,twocolumn,showpacs]{revtex4}

\usepackage[T1]{fontenc}
\usepackage{graphicx}
\usepackage{ae}

\begin{document}
\title{The role of wire imperfections in micro magnetic traps for atoms}
\author{J.~Estève$^{(1)}$, C.~Aussibal$^{(1)}$,
T.~Schumm$^{(1)}$, C.~Figl$^{(2)}$, D.~Mailly$^{(3)}$,
I.~Bouchoule$^{(1)}$, C.~I.~Westbrook$^{(1)}$ and
A.~Aspect$^{(1)}$}
\affiliation{
$^{(1)}$ Laboratoire Charles Fabry de l'Institut d'Optique, UMR 8501 du CNRS, 91403 Orsay, France \\
$^{(2)}$ Permanent address : Universit\"at Hannover, D 30167 Hannover, Germany\\
$^{(3)}$ Laboratoire de Photonique et de Nanostructures, UPR 20 du
CNRS, 91460 Marcoussis, France }

\begin{abstract}
We present a quantitative study of roughness in the magnitude of
the magnetic field produced by a current carrying microwire, {\it
i.e.} in the trapping potential for paramagnetic atoms. We show
that this potential roughness arises from deviations in the wire
current flow due to geometric fluctuations of the edges of the
wire : a measurement of the potential using cold trapped atoms
agrees with the potential computed from the measurement of the
wire edge roughness by a scanning electron microscope.
\end{abstract}

\pacs{39.25.+k, 03.75.Be}

\maketitle

The use of micro or even nano-fabricated electrical devices to
trap and manipulate cold atoms has attracted substantial interest,
especially since the demonstration of Bose-Einstein condensate
using such structures~\cite{ReichelNature,ZimmermannPRL}. Compact
and robust systems for producing BEC's, single mode waveguides and
possibly atom interferometers can now be envisaged. The small size
of the trapping elements, usually current-carrying wires producing
magnetic traps, and the proximity of the atoms to these elements
(typically tens of microns), means that many complex and rapidly
varying potentials can be designed~\cite{Folmanrevue}. This
approach has some disadvantages however. On the one hand it has
been shown that atoms are sensitive to the magnetic fields
generated by thermally fluctuating currents in a metal when they
are very
close~\cite{Folmanrevue,Zimmermann-fragPRA2002,Hinds-heating2003,Vuletic-VdW,Cornell-spinfliplosses}.
On the other hand, a time independent fragmentation of a cold
atomic cloud has been observed when atoms are brought close to a
current carrying
micro-wire~\cite{Zimmermann-fragPRA2002,Ketterle-guide2002,Hinds-frag2003}.

This fragmentation has been shown to be due to a potential
roughness arising from distortions of the current flow in the
wire~\cite{Zimmermann-frag2002,Ketterle-opttrap}. It has also been
demonstrated experimentally that the effect of these distortions
decreases with increasing distance from the
wire~\cite{Hinds-frag2003,Zimmermann-frag2002}. In an attempt to
account for the observations, a theoretical suggestion has been
made that the current distortions may be simply due to geometrical
deformations, more specifically meanders, of the
wire~\cite{Lukin-frag2003}. In this paper we show that for at
least one realization of a microfabricated magnetic trap, using
electroplating of gold, this suggestion is substantially correct.
We have measured the longitudinal density variation of a
fragmented thermal cloud of atoms trapped above a wire, and
inferred the rough magnetic potential. We have also made scanning
electron microscope images of the same wire and measured the
profile of the wire edges over the region explored by the atoms.
The magnetic potential as a function of position deduced from the
edge measurements is in good quantitative agreement with that
inferred from the atomic density. We suspect that this result is
not unique to our sample or fabrication process and we emphasize
the quantitative criterion for the necessary wire quality to be
used for atom manipulation.

\begin{figure}
\centerline{\includegraphics{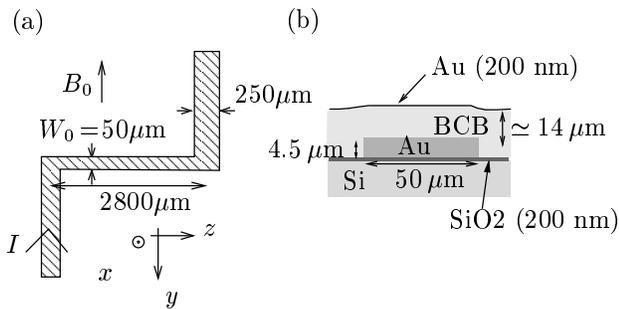}} \caption{ (a) : Z-wire
used to produce the magnetic trap. (b) : Cross section of the wire
in the ($xy$) plane. The wire is covered with a layer of BCB
polymer, and a thin gold layer acting as a mirror. The origin of
the coordinate system is taken at the center of the wire. }
\label{fig.puce}
\end{figure}

The wires we use are produced using standard microelectronic
techniques. A silicon wafer is first covered by a 200~nm silicon
dioxide layer. Next, layers of titanium (20~nm) and gold (200~nm)
are evaporated. The wire pattern is imprinted on a 6~$\mu$m thick
photoresist using optical lithography. Gold is electroplated
between the resist walls using the first gold layer as an
electrode. After removing the photoresist and the first gold and
titanium layers, we obtain electroplated wires of thickness
$u_0=4.5~\mu$m with a rectangular transverse profile (see
Fig.~\ref{fig.puce}). A planarizing dielectric layer (BCB, a
benzocyclobutene-based polymer) is deposited to cover the central
region of the chip. On top of the BCB, a 200~nm gold layer is
evaporated to be used as an optical mirror for light at 780~nm.
The distance between the center of the wire and the mirror layer
has been measured to be 14(1)~$\mu$m.

The magnetic trap is produced by a current $I$ flowing through a
Z-shaped microwire~\cite{Reic99} together with an external uniform
magnetic field $\bf{B_0}$ (along the $y$-axis, see
Fig.~\ref{fig.puce}) parallel to the chip surface and
perpendicular to the central part of the wire. The central part of
the Z-wire is 50~$\mu$m wide and 2800~$\mu$m long. Cold $^{87}$Rb
atoms, collected in a surface magneto-optical trap, are loaded
into the magnetic trap after a stage of optical molasses and
optical pumping to the $|F=2,m=2\rangle$ hyperfine state. The trap
is then compressed so that efficient forced evaporative cooling
can be applied. Finally, the trap is decompressed. Final values of
$I$ and $B_0$ vary from 200~mA to 300~mA and from 3~G to 14~G
respectively so that the height of the magnetic trap above the
wire ranges from 33\,$\mu$m to 170$\,\mu$m. An external
longitudinal magnetic field of a few Gauss aligned along $z$ is
added to limit the strength of the transverse confinement and to
avoid spin flip losses induced by technical noise. For these
parameters, the trap is highly elongated along the $z$-axis. The
transverse oscillation frequency is typically
$\omega_{\perp}/(2\pi)=3.5$\,kHz and 120\,Hz for traps at
33\,$\mu$m and 170\,$\mu$m from the wire respectively.

The potential roughness is deduced from measurements of the
longitudinal density distribution of cold trapped atoms. The
atomic density is probed using absorption imaging after the atoms
have been released from the final trap by switching off the
current in the Z-wire (switching time smaller than 100\,$\mu$s).
The probe beam is reflected by the chip at 45$^{\rm o}$ allowing
us to have two images of the cloud on the same picture. From
images taken just after (500\,$\mu$s) switching off the Z-wire
current we infer the longitudinal density $n(z)=\int\!\!\!\int\!
dxdy\, n(x,y,z)$. We also deduce the height of the atoms above the
mirror layer from the distance between the two images. The
temperature of the atoms is determined by measuring the expansion
of the cloud in the transverse direction after longer times of
flight (1 to 5\,ms).

To infer the longitudinal potential experienced by the atoms, we
assume the potential is given by
\begin{equation}
V(x,y,z) = V_z(z) + V_\mathrm{harm}(x,y)
\end{equation}
where $V_\mathrm{harm}(x,y)$ is a transverse harmonic potential.
Under this separability assumption, the longitudinal potential is
directly obtained from the measured longitudinal density of a
cloud at thermal equilibrium using the Boltzmann law $V_z(z) =-
k_{\rm B} T \ln(n(z))$. To maximize the sensitivity to the
longitudinal potential variations, we choose a temperature of the
same order as the variations ( $T\simeq 0.4~\mu$K for traps at
170\,$\mu$m from the wire and $T\simeq 2.2~\mu$K for traps at
33\,$\mu$m from the wire). The separability assumption has been
checked experimentally by deducing a $z$-dependent oscillation
frequency from the rms width of the transverse atomic density at
different positions. At 33\,$\mu$m from the wire, there is no
evidence of a varying oscillation frequency. At a height of
170~$\mu$m above the chip, over a longitudinal extent of
450~$\mu$m, we deduce a variation of the transverse oscillation
frequency of about 13\%. In this case, the assumption of
separability introduces an error of $0.2\,k_{\rm B}T$ in the
deduced potential.

The potential experienced by the atoms is $V=\mu_B |\vec{B}|$. To
a very good approximation, the magnetic field at the minimum of
$V_{\rm{harm}}$ is along the $z$-axis (for our parameters, the
deviation from the $z$-axis is computed to be always smaller than
1~mrad) so that the longitudinal potential is given by
$V_z(z)=\mu_B B_z(z)$. For a perfect Z-wire, the longitudinal
potential is solely due to the arms of the wire and has a smooth
shape. However, we observe a rough potential which is a signature
for the presence of an additional spatially fluctuating
longitudinal magnetic field. A spatially fluctuating transverse
magnetic field of similar amplitude would give rise to transverse
displacement of the potential which is undetectable with our
imaging resolution and small enough to leave our analysis
unchanged.

In the following, we present the data analysis which enables us to
extract the longitudinal potential roughness. In order to have a
large statistical sample and to gain access to low spatial
frequencies, one must measure the potential roughness over a large
fraction of the central wire. In our experiment, however, the
longitudinal confinement produced by the arms of the Z-wire itself
is too strong to enable the atomic cloud to spread over the full
extent of the central wire. To circumvent this difficulty, we add
an adjustable longitudinal gradient of $B_z$ which shifts the
atomic cloud along the central wire. We then measure the potential
above different zones of the central wire. We typically use four
different spatial zones which overlap each other by about
$200\,\mu$m. We then reconstruct the potential over the total
explored region by subtracting gradients from the potentials
obtained in the different zones. Those gradients are chosen in
order to minimize the difference between the potentials in the
regions where they overlap.

We are interested in those potential variations which differ from
the smooth confining potential due to the arms of the wire. We
thus subtract the expected confining potential of an ideal wire
from the reconstructed potential. To find the expected potential,
we model the arms of the Z by two infinitesimally thin,
semi-infinite wires of width 250\,$\mu$m, separated by a distance
$l$ and assume a uniform current distribution. We fit each
reconstructed potential to the sum of the result of the model and
a gradient, using the gradient, $l$ and the distance $h$ above the
wire as fitting parameters. The fitted values of $l$ differ by a
few percent from the nominal value 2.8~mm, while we find
$h=13(1)\,\mu$m$+d$ where $d$ is the distance from the mirror as
measured in the trap images. This result is consistent with the
measured 14$\,\mu$m thickness of the BCB layer.

\begin{figure}
{\includegraphics{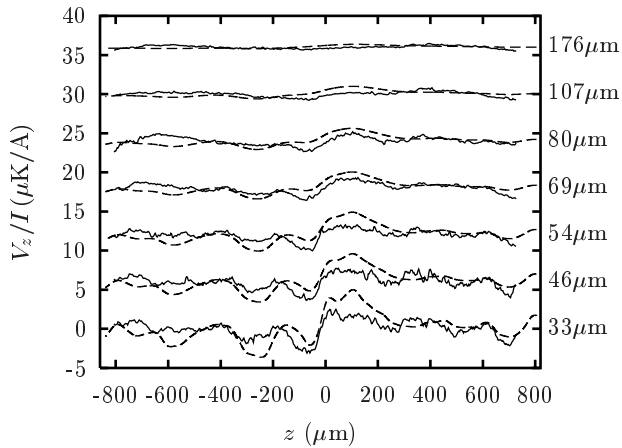}} \caption{ Rough potentials
normalized to the current in the Z-shaped wire for different
heights from the wire. Solid lines : potentials measured using
cold atomic clouds. Dashed lines : potentials calculated from the
measured geometric roughness of the edges of the wire. The
different curves have been shifted by $6\,\mu$K/A from each
other.} \label{fig.potentiel}
\end{figure}

\begin{figure}
\centerline{\includegraphics{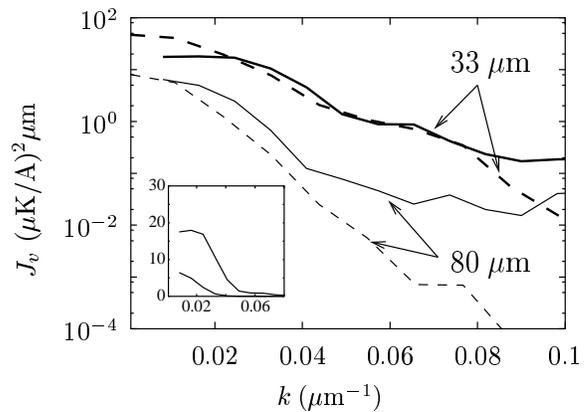}} \caption{
Potential spectral density $J_v=1/(2\pi
I^2)\int_{-\infty}^{\infty} \langle V_z(z)V_z(z+u)\rangle
e^{iku}du$ at $33\,\mu$m from the wire (fat lines) and at
$80\,\mu$m from the wire (thin lines). Solid lines : potential
measured using cold atomic clouds, the inset shows the curves on a
linear scale. Dashed lines : potentials calculated from the
measured geometric fluctuations of the edges of the wire. These
estimations of the spectral density are made with the Welch
algorithm~\cite{welch} using windows half the size of the total
explored region (1.6\,mm).
 }
\label{fig.compPwelch}
\end{figure}

The potential which remains after the above subtraction procedure
is plotted for different heights in Fig.~\ref{fig.potentiel}. For
a fixed trap height $h$ (fixed ratio $I/B_0$), we have checked
that the potential is proportional to the current in the wire;
therefore we normalize all measurements to the wire current. The
most obvious observation is that the amplitude of the roughness
decreases as one gets further away from the wire. The spectral
density of the potential roughness is shown in
Fig.~\ref{fig.compPwelch} for two different heights above the
wire. We observe that the spectrum gets narrower as the distance
from the wire increases (see inset in Fig.~\ref{fig.compPwelch}).
This is expected since fluctuations of wavelength much smaller
than the height above the wire are averaged to zero. At high wave
vectors ($k > 0.07\,\mu$m$^{-1}$ at 33\,$\mu$m and $k >
0.04\,\mu$m$^{-1}$ at 80\,$\mu$m) the spectrum exhibits plateaus
which we interpret as instrumental noise. We expect this noise
level to depend on our experimental parameters such as
temperature, current and atomic density. Qualitatively, smaller
atom-wire distances, which are analyzed with higher temperatures
should result in higher plateaus. This is consistent with the
observation.

\begin{figure}
\hspace*{0cm}{\includegraphics{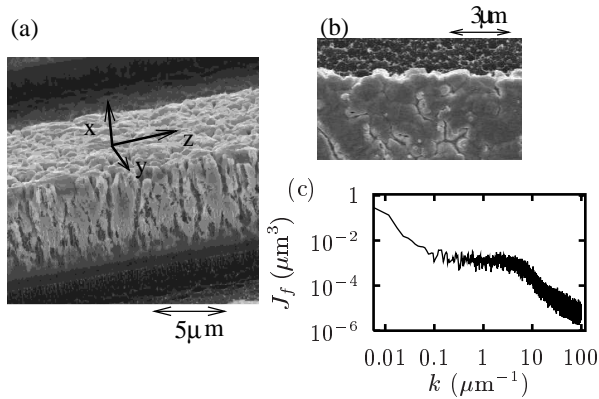}} \caption{
Imperfections of the edges of the wire. In the SEM image of the
wire taken from the side (a), one can see that the edge deviation
function $f$ is roughly independent of $x$. (c) Spectral density
of $f$ extracted from SEM images taken from the top as in Fig.(b).
} \label{fig.imagebord}
\end{figure}

In the following we evaluate the rough potential due to edge
fluctuations of the central wire. For this purpose, the edges of
the wire are imaged using a scanning electron microscope (SEM),
after removal of the BCB layer by reactive ion etching.
Figure~\ref{fig.imagebord}(a) indicates that the function $f$,
which gives the deviation of the position of the wire edge from
the mean position $y=\pm W_0/2$, is roughly independent of $x$. We
make the approximation that $f$ depends only on $z$. We deduce $f$
from SEM images taken from above the wire (see
Fig.~\ref{fig.imagebord}(b)). To resolve $f$, whose rms amplitude
is only $0.2\,\mu$m, we use fields of view as small as 50\,$\mu$m.
The function $f$ is reconstructed over the entire length of the
central wire using many images having about 18\,$\mu$m overlaps.
As shown in Fig.~\ref{fig.imagebord}(c), several length scales
appear in the spectrum. There are small fluctuations of
correlation length of about 100\,nm and, more importantly,
fluctuations of larger wavelength (60 to 1000\,$\mu$m).

The geometric fluctuations of the edges of the wire induce a
distortion of the current flow which produces a longitudinal
magnetic field roughness responsible for a potential roughness. To
compute the current density in the wire, we assume a uniform
resistivity inside the wire. We also assume that fluctuations are
small enough to make the current density distortion linear in
$f_{L/R}$, where $f_{L/R}$ are the fluctuations of the left and
right edge of the wire respectively. The current density is in the
$yz$-plane and, because the rough potential is proportional to the
longitudinal magnetic field, we are only interested in its $y$
component, $j_y$. Because of symmetry, only the part of $j_y(z,y)$
which is even in $y$ contributes to $B_z$ in the $xz$-plane. Thus,
only the symmetric component $f^+=1/2(f_L+f_R)$ is considered. The
Fourier component $f^+(k)$ of $f^+$ induces a transverse current
density~\cite{Lukin-frag2003}
\begin{equation}
j_y^+(k,y)=ik \, f^+(k) \frac{I}{W_0 \, u_0} \frac{{\rm
cosh}(ky)}{{\rm cosh}(kW_0/2)} . \label{eq.sigmaxk}
\end{equation}
As the distances from the wire we consider (33 to 176\,$\mu$m) are
much larger than the thickness of the wire $u_0=4.5\,\mu$m we will
in the following assume an infinitely flat wire. To efficiently
compute the longitudinal magnetic field produced by these current
distortions, we use the expansion on the modified Bessel functions
of second kind $K_n(kx)$, which is valid for $x>W_0 /2 $. This
expansion is, in the $xz$-plane,
\begin{equation}
B_{z}(k,x)=-\sum_{n\geq 0} k(c_{2n}(k)+c_{2n+2}(k)) K_{2n+1}(kx)
\label{eq.Bzflat}
\end{equation}
where
\begin{equation}
c_{2n}(k)=(-1)^{n}\frac{\mu_0}{\pi}u_0\int_0^{W_0/2} I_{2n}(ky)\
j_y^+(k,y) \ dy, \label{eq.cln}
\end{equation}
$I_n$ being the modified Bessel function of the first kind. For
small wave numbers $k$ such that $kW_0\ll 1$, one expects the
$c_n$ coefficients to decrease rapidly with $n$. Indeed, for $k
r\ll 1$, $I_n(kr) \simeq (kr)^n/(2^n n!)$. In our data analysis,
only $k$ wave vectors smaller than $0.07\,\mu$m$^{-1}$ are
considered for which $kW_0/2<0.8$ so we expect the $c_n$
coefficients to decrease rapidly with $n$. In the calculations,
only the terms up to $n=20$ are used.

Equations~\ref{eq.sigmaxk},\ref{eq.cln} and \ref{eq.Bzflat} then
allow us to compute the fluctuating potential from the measured
function $f$. In Fig.~\ref{fig.compPwelch}, we plot the spectral
density of the potential roughness calculated from $f$ for two
different heights above the wire (33 and 80\,$\mu$m) and compare
it with those obtained from the potential measured with the atoms.
For both heights, and for wave vectors small enough so that the
measurements made with the atoms are not limited by experimental
noise, the two curves are in good agreement. As we have measured
the $f$ function on the whole region explored by the atoms, we can
compute directly the expected potential roughness. In
Fig.~\ref{fig.potentiel}, this calculated potential roughness is
compared with the roughness measured with the atoms for different
heights above the wire. Remarkably the potential computed from the
wire edges and the one deduced from the atomic distributions have
not only consistent spectra but present well correlated profiles.
We thus conclude that the potential roughness is due to the
geometric fluctuations of the edges of the wire. The good
agreement between the curves also validates the assumption of
uniform conductivity inside the wire used to compute the current
distortion flow.

In conclusion, we have shown that the potential roughness we
observe can be attributed to the geometric fluctuations of the
wire edges. Fluctuations at low wave vectors, responsible for most
of the potential roughness, correspond to wire edge fluctuations
of very small amplitude compared to their correlation length. We
emphasize that a quantitative evaluation of these wire roughness
components demands dedicated measurement methods.

Furthermore, wire edge fluctuations put a lower limit on the
possibility of down-scaling atom chips. For a given fabrication
technology the wire edge fluctuations are expected to be
independent of the wire width $W_0$. Thus assuming a white noise
spectrum, the normalized potential roughness $V_{\rm rms}/I$
varies as $1/W_0^{5/2}$ for fixed ratio $h/W_0$, $h$ being the
distance to the wire~\cite{Lukin-frag2003}. In order to reduce the
potential roughness, one must pay careful attention to edge
fluctuations when choosing a fabrication process. For example, we
are currently investigating electron beam lithography followed by
gold evaporation. Preliminary measurements indicate a reduction of
the spectral density of the wire edge fluctuations by at least two
orders of magnitude for wave vectors ranging from
0.1~$\mu$m$^{-1}$ to 10~$\mu$m$^{-1}$. This should allow us to
reduce the spectral density of the potential roughness by the same
factor unless a new as yet unobserved phenomenon such as bulk
inhomogeneity sets a new limit on atom chip
down-scaling~\cite{Schmiedmayer}.

We thank C. Henkel and H. Nguyen for helpful discussions. This
work was supported by E.U. (IST-2001-38863, HPRN-CT-2000-00125),
by DGA (03.34.033) and by the French ministry of research (Action
concertée "nanosciences-nanotechnologies").

\end{document}